\documentclass[rnaas]{aastex631}

\newcommand{\Msun}{\mbox{$\rm M_{\odot}$}}

\newcommand{\e}[1]{\times 10^{#1}}
\graphicspath{{./}{figures/}}
\shortauthors{Sepulveda et al.}
\submitjournal{Research Notes of the American Astronomical Society}
\begin{document}
\title{The Hybrid Debris Disk Host Star HD 21997 is a High-Frequency Delta Scuti Pulsator}
\correspondingauthor{Aldo G. Sepulveda}
\email{aldo.sepulveda@hawaii.edu}
\author[0000-0002-8621-2682]{Aldo G. Sepulveda}
\altaffiliation{NSF Graduate Research Fellow}
\affiliation{Institute for Astronomy, University of Hawai`i at M\={a}noa. 2680 Woodlawn Drive, Honolulu, HI 96822, USA}

\author[0000-0001-5222-4661]{Timothy R. Bedding}
\affiliation{Sydney Institute for Astronomy (SIfA), School of Physics, University of Sydney, NSW 2006, Australia}

\author[0000-0002-5648-3107]{Simon J. Murphy}
\affiliation{Centre for Astrophysics, University of Southern Queensland, Toowoomba, QLD 4350, Australia}

\author[0000-0003-4705-3188]{Luca Matr\`a}
\affiliation{School of Physics, Trinity College Dublin, College Green, Dublin 2, Ireland}

\author[0000-0001-8832-4488]{Daniel Huber}
\affiliation{Institute for Astronomy, University of Hawai`i at M\={a}noa. 2680 Woodlawn Drive, Honolulu, HI 96822, USA}
\affiliation{Sydney Institute for Astronomy (SIfA), School of Physics, University of Sydney, NSW 2006, Australia}

\author[0000-0002-3726-4881]{Zhoujian Zhang}
\altaffiliation{NASA Sagan Fellow}
\affiliation{Department of Astronomy and Astrophysics, University of California, Santa Cruz, CA 95064, USA}

\begin{abstract}
HD~21997 is host to a prototypical ``hybrid" debris disk characterized by debris disk-like dust properties and a CO gas mass comparable to a protoplanetary disk. We use Transiting Exoplanet Survey Satellite time series photometry to demonstrate that HD~21997 is a high-frequency delta Scuti pulsator. If the mode identification can be unambiguously determined in future works, an asteroseismic age of HD~21997 may become feasible.
\end{abstract}

\keywords{Delta Scuti variable stars (370) --- Stellar pulsations (1625) --- Variable stars (1761)} 
\section{Introduction}
HD~21997, a $\sim$1.8\Msun\ A3 star, hosts a debris disk bearing a CO gas mass that is large for its age and dust properties \citep{Kospal+2013,Moor+2013}. Although the presence of gas in debris disks is now known in over 20 systems, for a subset of these, including HD~21997, it is so large that it challenges models of second-generation exocometary gas release. These have been termed ``hybrid" debris disks due to the potential coexistence of second-generation dust with primordial gas \citep{Kospal+2013,Kral+2019}. Accurate age constraints for HD~21997 are necessary to interpret the dust and gas evolution both for the system and for hybrid disks as a population. In this context, asteroseismic modeling of intermediate-mass delta Scuti ($\delta$~Scuti) pulsators has the potential to yield a precise age \citep[e.g.,][]{Murphy+2021}. We therefore investigate whether HD~21997 is a $\delta$~Scuti pulsator.
\section{Pulsation Classification \& Discussion}
\begin{figure*}[ht!]

  \includegraphics[width=7.1in]{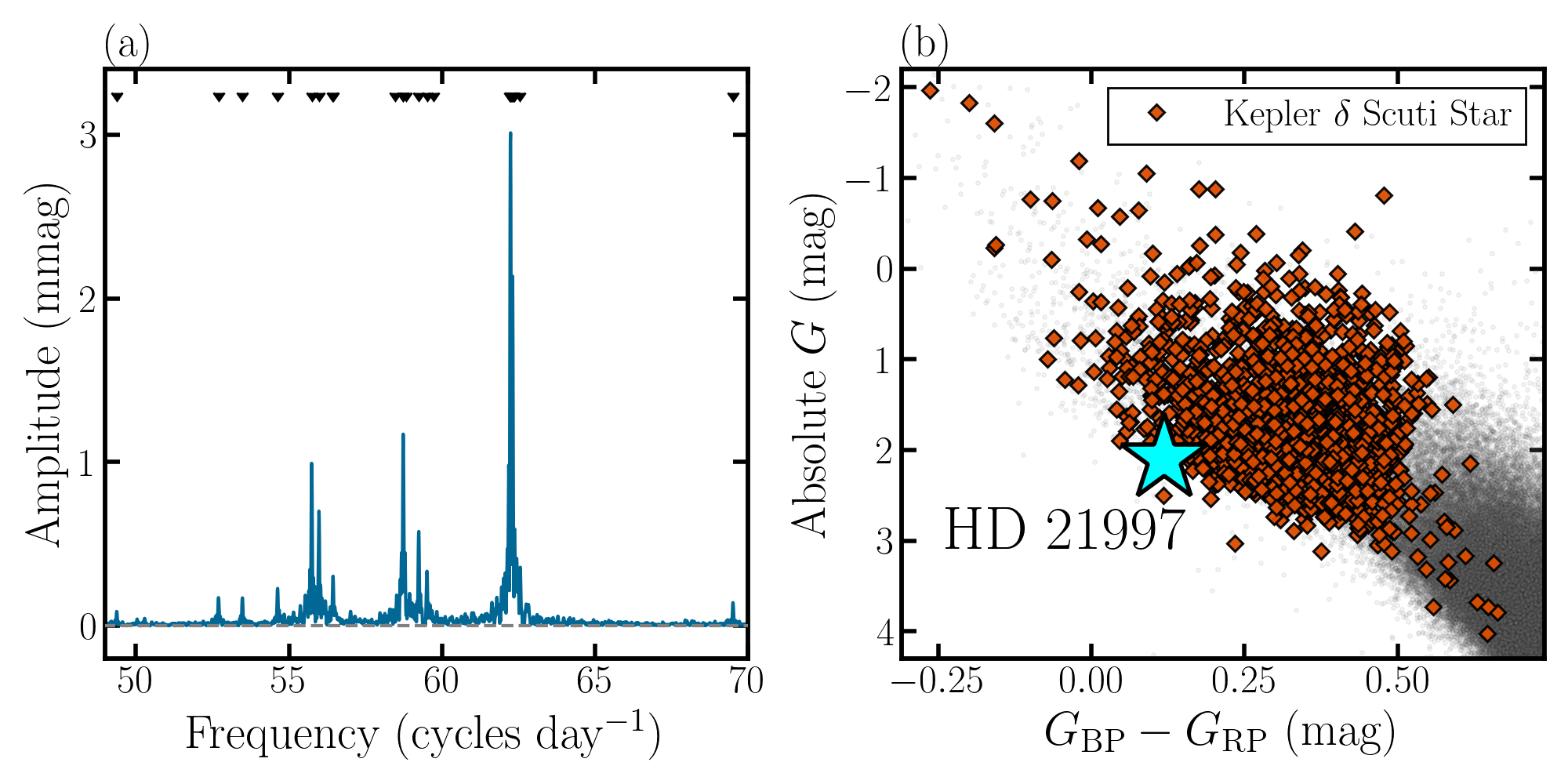}
  \centering
  \caption{(a): TESS amplitude spectrum of HD~21997 zoomed in on the region of significant pulsations. Significant frequencies from our prewhitening procedure are denoted by black downward triangles. (b): \textit{Gaia} DR3 color-magnitude diagram generated following \citet{Sepulveda+2022}. Orange diamonds are a sample of \textit{Kepler} $\delta$~Scuti stars from \citet{Murphy+2019} and small grey dots are non-pulsating KIC stars. HD~21997, whose color and absolute magnitude were calculated using $A_V = 0.075$ \citep{Chen+2014}, is overplotted as a large cyan star. }
    \label{fig:DS}
\end{figure*}

We downloaded the Transiting Exoplanet Survey Satellite (TESS, \citealt{TESSMission}) two-minute cadence PDC-SAP SPOC light curve \citep{smith12,stumpe12,stumpe14,jenkins16} of HD~21997 observed in Sector 4 (2018 Oct 19 --Nov 14). In the resulting amplitude spectrum, we identify 25 significant frequencies spanning $\sim$49--70 cycles~day$^{-1}$ (Figure \ref{fig:DS}a) by iteratively fitting sine waves to the time series down to a spectral significance threshold of 20 using \texttt{Sigspec} \citep{SigSpec}. These frequencies are consistent with pressure modes of $\delta$~Scuti stars. Following the procedure described in \citet{Sepulveda+2022}, in Figure \ref{fig:DS}b we compare the Gaia DR3 $BP-RP$ color and absolute $G$ magnitude of HD 21997 \citep[][]{GaiaMission,GaiaEDR3} to that of the population of Kepler $\delta$~Scuti stars from \citet{Murphy+2019}.

\newpage

A precursor to asteroseismic modeling of HD~21997 is an accurate mode identification. By inspecting \'{e}chelle diagrams, we find plausible large-frequency separation ($\Delta \nu$) values in the range of 6.13--6.56~cycles~day$^{-1}$, which are all consistent within the range found in \citet{Bedding+2020}. However, the ambiguity in $\Delta \nu$ precludes a straightforward mode identification.

Our detection of high-frequency $\delta$~Scuti pulsations is independent qualitative evidence for HD~21997's youth \citep{Bedding+2020}, which is furthermore supported by its position in our Gaia DR3 color-magnitude diagram. \citet{Ujjwal+2020} used isochrone fitting together with Gaia DR2 data of HD~21997 to estimate an age of 38.5~Myr. HD~21997 is an associated member of the Columba young moving group \citep{Torres+2008}, with age estimates spanning 20--40 Myr  \citep[e.g.,][]{Torres+2008,Bell+2015,Ujjwal+2020}. These properties support that HD~21997 is likely $<$50~Myr. Yet, how accurately these ages represent HD~21997 is subject to further inquiry. The isochrone age estimate from \citet{Ujjwal+2020} has no reported uncertainty, which would include model systematics from the isochrones used in addition to any statistical uncertainty. In addition, the rotation period of HD~21997 is unknown, and may have a large effect on the star's position on the HR diagram, and therefore also with respect to any isochrones \citep{PerezHernandez+1999}. Moving group members themselves generally have age spreads with respect to the common age adopted for the group, and furthermore moving group ages themselves are subject to systematics of which stars and methods were used for the age constraints. If the asteroseismic mode identification can be achieved for HD~21997 in the future, an asteroseismic age of 10--20\% precision may be feasible. 

\begin{acknowledgments}

\textit{Acknowledgments}. A.G.S. acknowledges support from the NSF GRFP (Grant No. 1842402 and 2236415). L.M. acknowledges support by the Irish Research Council under grant IRCLA/2022/ 3788. This paper includes data collected by the TESS mission, which are publicly available from MAST. Funding for the TESS mission is provided by the NASA’s Science Mission Directorate.
\end{acknowledgments}

\facilities{TESS}
\software{\texttt{lightkurve} \citep{LightkurveCollaboration+2018}, \texttt{echelle} \citep{Hey+Ball2020}, \texttt{SigSpec} \citep{SigSpec}, \texttt{matplotlib} \citep{matplotlib}, \texttt{numpy} \citep{Harris+2020}, \texttt{astroquery} \citep{astroquery},}
\bibliographystyle{aasjournal}
\bibliography{ms}

\begin{thebibliography}{}
\expandafter\ifx\csname natexlab\endcsname\relax\def\natexlab#1{#1}\fi
\providecommand{\url}[1]{\href{#1}{#1}}
\providecommand{\dodoi}[1]{doi:~\href{http://doi.org/#1}{\nolinkurl{#1}}}
\providecommand{\doeprint}[1]{\href{http://ascl.net/#1}{\nolinkurl{http://ascl.net/#1}}}
\providecommand{\doarXiv}[1]{\href{https://arxiv.org/abs/#1}{\nolinkurl{https://arxiv.org/abs/#1}}}

\bibitem[{{Bedding} {et~al.}(2020){Bedding}, {Murphy}, {Hey}, {Huber}, {Li}, {Smalley}, {Stello}, {White}, {Ball}, {Chaplin}, {Colman}, {Fuller}, {Gaidos}, {Harbeck}, {Hermes}, {Holdsworth}, {Li}, {Li}, {Mann}, {Reese}, {Sekaran}, {Yu}, {Antoci}, {Bergmann}, {Brown}, {Howard}, {Ireland}, {Isaacson}, {Jenkins}, {Kjeldsen}, {McCully}, {Rabus}, {Rains}, {Ricker}, {Tinney}, \& {Vanderspek}}]{Bedding+2020}
{Bedding}, T.~R., {Murphy}, S.~J., {Hey}, D.~R., {et~al.} 2020, \nat, 581, 147, \dodoi{10.1038/s41586-020-2226-8}

\bibitem[{{Bell} {et~al.}(2015){Bell}, {Mamajek}, \& {Naylor}}]{Bell+2015}
{Bell}, C. P.~M., {Mamajek}, E.~E., \& {Naylor}, T. 2015, \mnras, 454, 593, \dodoi{10.1093/mnras/stv1981}

\bibitem[{{Chen} {et~al.}(2014){Chen}, {Mittal}, {Kuchner}, {Forrest}, {Lisse}, {Manoj}, {Sargent}, \& {Watson}}]{Chen+2014}
{Chen}, C.~H., {Mittal}, T., {Kuchner}, M., {et~al.} 2014, \apjs, 211, 25, \dodoi{10.1088/0067-0049/211/2/25}

\bibitem[{{Gaia Collaboration} {et~al.}(2016){Gaia Collaboration}, {Prusti}, {de Bruijne}, {Brown}, {Vallenari}, {Babusiaux}, {Bailer-Jones}, {Bastian}, {Biermann}, {Evans}, \& et~al.}]{GaiaMission}
{Gaia Collaboration}, {Prusti}, T., {de Bruijne}, J.~H.~J., {et~al.} 2016, \aap, 595, A1, \dodoi{10.1051/0004-6361/201629272}

\bibitem[{{Gaia Collaboration} {et~al.}(2021){Gaia Collaboration}, {Brown}, {Vallenari}, {Prusti}, {de Bruijne}, {Babusiaux}, {Biermann}, {Creevey}, {Evans}, {Eyer}, {Hutton}, {Jansen}, {Jordi}, {Klioner}, {Lammers}, {Lindegren}, {Luri}, {Mignard}, {Panem}, {Pourbaix}, {Randich}, {Sartoretti}, {Soubiran}, {Walton}, {Arenou}, {Bailer-Jones}, {Bastian}, {Cropper}, {Drimmel}, {Katz}, {Lattanzi}, {van Leeuwen}, {Bakker}, {Cacciari}, {Casta{\~n}eda}, {De Angeli}, {Ducourant}, {Fabricius}, {Fouesneau}, {Fr{\'e}mat}, {Guerra}, {Guerrier}, {Guiraud}, {Jean-Antoine Piccolo}, {Masana}, {Messineo}, {Mowlavi}, {Nicolas}, {Nienartowicz}, {Pailler}, {Panuzzo}, {Riclet}, {Roux}, {Seabroke}, {Sordo}, {Tanga}, {Th{\'e}venin}, {Gracia-Abril}, {Portell}, {Teyssier}, {Altmann}, {Andrae}, {Bellas-Velidis}, {Benson}, {Berthier}, {Blomme}, {Brugaletta}, {Burgess}, {Busso}, {Carry}, {Cellino}, {Cheek}, {Clementini}, {Damerdji}, {Davidson}, {Delchambre}, {Dell'Oro}, {Fern{\'a}ndez-Hern{\'a}ndez}, {Galluccio}, {Garc{\'\i}a-Lario},
  {Garcia-Reinaldos}, {Gonz{\'a}lez-N{\'u}{\~n}ez}, {Gosset}, {Haigron}, {Halbwachs}, {Hambly}, {Harrison}, {Hatzidimitriou}, {Heiter}, {Hern{\'a}ndez}, {Hestroffer}, {Hodgkin}, {Holl}, {Jan{\ss}en}, {Jevardat de Fombelle}, {Jordan}, {Krone-Martins}, {Lanzafame}, {L{\"o}ffler}, {Lorca}, {Manteiga}, {Marchal}, {Marrese}, {Moitinho}, {Mora}, {Muinonen}, {Osborne}, {Pancino}, {Pauwels}, {Petit}, {Recio-Blanco}, {Richards}, {Riello}, {Rimoldini}, {Robin}, {Roegiers}, {Rybizki}, {Sarro}, {Siopis}, {Smith}, {Sozzetti}, {Ulla}, {Utrilla}, {van Leeuwen}, {van Reeven}, {Abbas}, {Abreu Aramburu}, {Accart}, {Aerts}, {Aguado}, {Ajaj}, {Altavilla}, {{\'A}lvarez}, {{\'A}lvarez Cid-Fuentes}, {Alves}, {Anderson}, {Anglada Varela}, {Antoja}, {Audard}, {Baines}, {Baker}, {Balaguer-N{\'u}{\~n}ez}, {Balbinot}, {Balog}, {Barache}, {Barbato}, {Barros}, {Barstow}, {Bartolom{\'e}}, {Bassilana}, {Bauchet}, {Baudesson-Stella}, {Becciani}, {Bellazzini}, {Bernet}, {Bertone}, {Bianchi}, {Blanco-Cuaresma}, {Boch}, {Bombrun}, {Bossini},
  {Bouquillon}, {Bragaglia}, {Bramante}, {Breedt}, {Bressan}, {Brouillet}, {Bucciarelli}, {Burlacu}, {Busonero}, {Butkevich}, {Buzzi}, {Caffau}, {Cancelliere}, {C{\'a}novas}, {Cantat-Gaudin}, {Carballo}, {Carlucci}, {Carnerero}, {Carrasco}, {Casamiquela}, {Castellani}, {Castro-Ginard}, {Castro Sampol}, {Chaoul}, {Charlot}, {Chemin}, {Chiavassa}, {Cioni}, {Comoretto}, {Cooper}, {Cornez}, {Cowell}, {Crifo}, {Crosta}, {Crowley}, {Dafonte}, {Dapergolas}, {David}, {David}, {de Laverny}, {De Luise}, {De March}, {De Ridder}, {de Souza}, {de Teodoro}, {de Torres}, {del Peloso}, {del Pozo}, {Delbo}, {Delgado}, {Delgado}, {Delisle}, {Di Matteo}, {Diakite}, {Diener}, {Distefano}, {Dolding}, {Eappachen}, {Edvardsson}, {Enke}, {Esquej}, {Fabre}, {Fabrizio}, {Faigler}, {Fedorets}, {Fernique}, {Fienga}, {Figueras}, {Fouron}, {Fragkoudi}, {Fraile}, {Franke}, {Gai}, {Garabato}, {Garcia-Gutierrez}, {Garc{\'\i}a-Torres}, {Garofalo}, {Gavras}, {Gerlach}, {Geyer}, {Giacobbe}, {Gilmore}, {Girona}, {Giuffrida}, {Gomel}, {Gomez},
  {Gonzalez-Santamaria}, {Gonz{\'a}lez-Vidal}, {Granvik}, {Guti{\'e}rrez-S{\'a}nchez}, {Guy}, {Hauser}, {Haywood}, {Helmi}, {Hidalgo}, {Hilger}, {H{\l}adczuk}, {Hobbs}, {Holland}, {Huckle}, {Jasniewicz}, {Jonker}, {Juaristi Campillo}, {Julbe}, {Karbevska}, {Kervella}, {Khanna}, {Kochoska}, {Kontizas}, {Kordopatis}, {Korn}, {Kostrzewa-Rutkowska}, {Kruszy{\'n}ska}, {Lambert}, {Lanza}, {Lasne}, {Le Campion}, {Le Fustec}, {Lebreton}, {Lebzelter}, {Leccia}, {Leclerc}, {Lecoeur-Taibi}, {Liao}, {Licata}, {Lindstr{\o}m}, {Lister}, {Livanou}, {Lobel}, {Madrero Pardo}, {Managau}, {Mann}, {Marchant}, {Marconi}, {Marcos Santos}, {Marinoni}, {Marocco}, {Marshall}, {Martin Polo}, {Mart{\'\i}n-Fleitas}, {Masip}, {Massari}, {Mastrobuono-Battisti}, {Mazeh}, {McMillan}, {Messina}, {Michalik}, {Millar}, {Mints}, {Molina}, {Molinaro}, {Moln{\'a}r}, {Montegriffo}, {Mor}, {Morbidelli}, {Morel}, {Morris}, {Mulone}, {Munoz}, {Muraveva}, {Murphy}, {Musella}, {Noval}, {Ord{\'e}novic}, {Orr{\`u}}, {Osinde}, {Pagani}, {Pagano},
  {Palaversa}, {Palicio}, {Panahi}, {Pawlak}, {Pe{\~n}alosa Esteller}, {Penttil{\"a}}, {Piersimoni}, {Pineau}, {Plachy}, {Plum}, {Poggio}, {Poretti}, {Poujoulet}, {Pr{\v{s}}a}, {Pulone}, {Racero}, {Ragaini}, {Rainer}, {Raiteri}, {Rambaux}, {Ramos}, {Ramos-Lerate}, {Re Fiorentin}, {Regibo}, {Reyl{\'e}}, {Ripepi}, {Riva}, {Rixon}, {Robichon}, {Robin}, {Roelens}, {Rohrbasser}, {Romero-G{\'o}mez}, {Rowell}, {Royer}, {Rybicki}, {Sadowski}, {Sagrist{\`a} Sell{\'e}s}, {Sahlmann}, {Salgado}, {Salguero}, {Samaras}, {Sanchez Gimenez}, {Sanna}, {Santove{\~n}a}, {Sarasso}, {Schultheis}, {Sciacca}, {Segol}, {Segovia}, {S{\'e}gransan}, {Semeux}, {Shahaf}, {Siddiqui}, {Siebert}, {Siltala}, {Slezak}, {Smart}, {Solano}, {Solitro}, {Souami}, {Souchay}, {Spagna}, {Spoto}, {Steele}, {Steidelm{\"u}ller}, {Stephenson}, {S{\"u}veges}, {Szabados}, {Szegedi-Elek}, {Taris}, {Tauran}, {Taylor}, {Teixeira}, {Thuillot}, {Tonello}, {Torra}, {Torra}, {Turon}, {Unger}, {Vaillant}, {van Dillen}, {Vanel}, {Vecchiato}, {Viala}, {Vicente},
  {Voutsinas}, {Weiler}, {Wevers}, {Wyrzykowski}, {Yoldas}, {Yvard}, {Zhao}, {Zorec}, {Zucker}, {Zurbach}, \& {Zwitter}}]{GaiaEDR3}
{Gaia Collaboration}, {Brown}, A.~G.~A., {Vallenari}, A., {et~al.} 2021, \aap, 649, A1, \dodoi{10.1051/0004-6361/202039657}

\bibitem[{{Ginsburg} {et~al.}(2019){Ginsburg}, {Sip{\H{o}}cz}, {Brasseur}, {Cowperthwaite}, {Craig}, {Deil}, {Guillochon}, {Guzman}, {Liedtke}, {Lian Lim}, {Lockhart}, {Mommert}, {Morris}, {Norman}, {Parikh}, {Persson}, {Robitaille}, {Segovia}, {Singer}, {Tollerud}, {de Val-Borro}, {Valtchanov}, {Woillez}, {Astroquery Collaboration}, \& {a subset of astropy Collaboration}}]{astroquery}
{Ginsburg}, A., {Sip{\H{o}}cz}, B.~M., {Brasseur}, C.~E., {et~al.} 2019, \aj, 157, 98, \dodoi{10.3847/1538-3881/aafc33}

\bibitem[{{Harris} {et~al.}(2020){Harris}, {Millman}, {van der Walt}, {Gommers}, {Virtanen}, {Cournapeau}, {Wieser}, {Taylor}, {Berg}, {Smith}, {Kern}, {Picus}, {Hoyer}, {van Kerkwijk}, {Brett}, {Haldane}, {del R{\'\i}o}, {Wiebe}, {Peterson}, {G{\'e}rard-Marchant}, {Sheppard}, {Reddy}, {Weckesser}, {Abbasi}, {Gohlke}, \& {Oliphant}}]{Harris+2020}
{Harris}, C.~R., {Millman}, K.~J., {van der Walt}, S.~J., {et~al.} 2020, \nat, 585, 357, \dodoi{10.1038/s41586-020-2649-2}

\bibitem[{Hey \& Ball(2020)}]{Hey+Ball2020}
Hey, D., \& Ball, W. 2020, {Echelle: Dynamic echelle diagrams for asteroseismology}, 1.4,  Zenodo, \dodoi{10.5281/zenodo.3629933}

\bibitem[{Hunter(2007)}]{matplotlib}
Hunter, J.~D. 2007, Computing in Science Engineering, 9, 90, \dodoi{10.1109/MCSE.2007.55}

\bibitem[{{Jenkins} {et~al.}(2016){Jenkins}, {Twicken}, {McCauliff}, {Campbell}, {Sanderfer}, {Lung}, {Mansouri-Samani}, {Girouard}, {Tenenbaum}, {Klaus}, {Smith}, {Caldwell}, {Chacon}, {Henze}, {Heiges}, {Latham}, {Morgan}, {Swade}, {Rinehart}, \& {Vanderspek}}]{jenkins16}
{Jenkins}, J.~M., {Twicken}, J.~D., {McCauliff}, S., {et~al.} 2016, in SPIE Conference Series, Vol. 9913, \procspie, 99133E, \dodoi{10.1117/12.2233418}

\bibitem[{{K{\'o}sp{\'a}l} {et~al.}(2013){K{\'o}sp{\'a}l}, {Mo{\'o}r}, {Juh{\'a}sz}, {{\'A}brah{\'a}m}, {Apai}, {Csengeri}, {Grady}, {Henning}, {Hughes}, {Kiss}, {Pascucci}, \& {Schmalzl}}]{Kospal+2013}
{K{\'o}sp{\'a}l}, {\'A}., {Mo{\'o}r}, A., {Juh{\'a}sz}, A., {et~al.} 2013, \apj, 776, 77, \dodoi{10.1088/0004-637X/776/2/77}

\bibitem[{{Kral} {et~al.}(2019){Kral}, {Marino}, {Wyatt}, {Kama}, \& {Matr{\`a}}}]{Kral+2019}
{Kral}, Q., {Marino}, S., {Wyatt}, M.~C., {Kama}, M., \& {Matr{\`a}}, L. 2019, \mnras, 489, 3670, \dodoi{10.1093/mnras/sty2923}

\bibitem[{{Lightkurve Collaboration} {et~al.}(2018){Lightkurve Collaboration}, {Cardoso}, {Hedges}, {Gully-Santiago}, {Saunders}, {Cody}, {Barclay}, {Hall}, {Sagear}, {Turtelboom}, {Zhang}, {Tzanidakis}, {Mighell}, {Coughlin}, {Bell}, {Berta-Thompson}, {Williams}, {Dotson}, \& {Barentsen}}]{LightkurveCollaboration+2018}
{Lightkurve Collaboration}, {Cardoso}, J.~V.~d.~M., {Hedges}, C., {et~al.} 2018, {Lightkurve: Kepler and TESS time series analysis in Python}, Astrophysics Source Code Library.
\newblock \doeprint{1812.013}

\bibitem[{{Mo{\'o}r} {et~al.}(2013){Mo{\'o}r}, {Juh{\'a}sz}, {K{\'o}sp{\'a}l}, {{\'A}brah{\'a}m}, {Apai}, {Csengeri}, {Grady}, {Henning}, {Hughes}, {Kiss}, {Pascucci}, {Schmalzl}, \& {Gab{\'a}nyi}}]{Moor+2013}
{Mo{\'o}r}, A., {Juh{\'a}sz}, A., {K{\'o}sp{\'a}l}, {\'A}., {et~al.} 2013, \apjl, 777, L25, \dodoi{10.1088/2041-8205/777/2/L25}

\bibitem[{{Murphy} {et~al.}(2019){Murphy}, {Hey}, {Van Reeth}, \& {Bedding}}]{Murphy+2019}
{Murphy}, S.~J., {Hey}, D., {Van Reeth}, T., \& {Bedding}, T.~R. 2019, \mnras, 485, 2380, \dodoi{10.1093/mnras/stz590}

\bibitem[{{Murphy} {et~al.}(2021){Murphy}, {Joyce}, {Bedding}, {White}, \& {Kama}}]{Murphy+2021}
{Murphy}, S.~J., {Joyce}, M., {Bedding}, T.~R., {White}, T.~R., \& {Kama}, M. 2021, \mnras, 502, 1633, \dodoi{10.1093/mnras/stab144}

\bibitem[{{P{\'e}rez Hern{\'a}ndez} {et~al.}(1999){P{\'e}rez Hern{\'a}ndez}, {Claret}, {Hern{\'a}ndez}, \& {Michel}}]{PerezHernandez+1999}
{P{\'e}rez Hern{\'a}ndez}, F., {Claret}, A., {Hern{\'a}ndez}, M.~M., \& {Michel}, E. 1999, \aap, 346, 586

\bibitem[{{Reegen}(2007)}]{SigSpec}
{Reegen}, P. 2007, \aap, 467, 1353, \dodoi{10.1051/0004-6361:20066597}

\bibitem[{{Ricker} {et~al.}(2015){Ricker}, {Winn}, {Vanderspek}, {Latham}, {Bakos}, {Bean}, {Berta-Thompson}, {Brown}, {Buchhave}, {Butler}, {Butler}, {Chaplin}, {Charbonneau}, {Christensen-Dalsgaard}, {Clampin}, {Deming}, {Doty}, {De Lee}, {Dressing}, {Dunham}, {Endl}, {Fressin}, {Ge}, {Henning}, {Holman}, {Howard}, {Ida}, {Jenkins}, {Jernigan}, {Johnson}, {Kaltenegger}, {Kawai}, {Kjeldsen}, {Laughlin}, {Levine}, {Lin}, {Lissauer}, {MacQueen}, {Marcy}, {McCullough}, {Morton}, {Narita}, {Paegert}, {Palle}, {Pepe}, {Pepper}, {Quirrenbach}, {Rinehart}, {Sasselov}, {Sato}, {Seager}, {Sozzetti}, {Stassun}, {Sullivan}, {Szentgyorgyi}, {Torres}, {Udry}, \& {Villasenor}}]{TESSMission}
{Ricker}, G.~R., {Winn}, J.~N., {Vanderspek}, R., {et~al.} 2015, Journal of Astronomical Telescopes, Instruments, and Systems, 1, 014003, \dodoi{10.1117/1.JATIS.1.1.014003}

\bibitem[{{Sepulveda} {et~al.}(2022){Sepulveda}, {Huber}, {Zhang}, {Li}, {Liu}, \& {Bedding}}]{Sepulveda+2022}
{Sepulveda}, A.~G., {Huber}, D., {Zhang}, Z., {et~al.} 2022, \apj, 938, 49, \dodoi{10.3847/1538-4357/ac9229}

\bibitem[{{Smith} {et~al.}(2012){Smith}, {Stumpe}, {Van Cleve}, {Jenkins}, {Barclay}, {Fanelli}, {Girouard}, {Kolodziejczak}, {McCauliff}, {Morris}, \& {Twicken}}]{smith12}
{Smith}, J.~C., {Stumpe}, M.~C., {Van Cleve}, J.~E., {et~al.} 2012, \pasp, 124, 1000, \dodoi{10.1086/667697}

\bibitem[{{Stumpe} {et~al.}(2014){Stumpe}, {Smith}, {Catanzarite}, {Van Cleve}, {Jenkins}, {Twicken}, \& {Girouard}}]{stumpe14}
{Stumpe}, M.~C., {Smith}, J.~C., {Catanzarite}, J.~H., {et~al.} 2014, \pasp, 126, 100, \dodoi{10.1086/674989}

\bibitem[{{Stumpe} {et~al.}(2012){Stumpe}, {Smith}, {Van Cleve}, {Twicken}, {Barclay}, {Fanelli}, {Girouard}, {Jenkins}, {Kolodziejczak}, {McCauliff}, \& {Morris}}]{stumpe12}
{Stumpe}, M.~C., {Smith}, J.~C., {Van Cleve}, J.~E., {et~al.} 2012, \pasp, 124, 985, \dodoi{10.1086/667698}

\bibitem[{{Torres} {et~al.}(2008){Torres}, {Quast}, {Melo}, \& {Sterzik}}]{Torres+2008}
{Torres}, C.~A.~O., {Quast}, G.~R., {Melo}, C.~H.~F., \& {Sterzik}, M.~F. 2008, {Young Nearby Loose Associations}, ed. B.~{Reipurth}, Vol.~5, 757

\bibitem[{{Ujjwal} {et~al.}(2020){Ujjwal}, {Kartha}, {Mathew}, {Manoj}, \& {Narang}}]{Ujjwal+2020}
{Ujjwal}, K., {Kartha}, S.~S., {Mathew}, B., {Manoj}, P., \& {Narang}, M. 2020, \aj, 159, 166, \dodoi{10.3847/1538-3881/ab76d6}

\end{thebibliography}

\end{document}